\documentclass[sigconf]{acmart}

\AtBeginDocument{%
  \providecommand\BibTeX{{%
    \normalfont B\kern-0.5em{\scshape i\kern-0.25em b}\kern-0.8em\TeX}}}

\copyrightyear{2021}
\acmYear{2021}
\setcopyright{acmcopyright}
\acmConference[CIKM '21]{Proceedings of the 30th ACM International Conference on Information and Knowledge Management}{November 1--5, 2021}{Virtual Event, QLD, Australia}
\acmBooktitle{Proceedings of the 30th ACM International Conference on Information and Knowledge Management (CIKM '21), November 1--5, 2021, Virtual Event, QLD, Australia}
\acmPrice{15.00}
\acmDOI{10.1145/3459637.3482312}
\acmISBN{978-1-4503-8446-9/21/11}

\usepackage{bm}

\usepackage[T1]{fontenc}
\usepackage{graphicx} 
\usepackage{subfigure}
\usepackage{makecell}
\usepackage{threeparttable}
\usepackage{multirow}
\usepackage{mathrsfs}
\usepackage{bbding}
\usepackage{pifont}

\usepackage{float}
\usepackage{bm}
\usepackage{array}
\usepackage{booktabs}
\usepackage{soul}
\usepackage{color}

\newcommand{\tabincell}[2]{\begin{tabular}{@{}#1@{}}#2\end{tabular}}
\usepackage{hyperref}
\acmSubmissionID{208}

\begin{document}
\fancyhead{}
\title{Zero Shot on the Cold-Start Problem: \\Model-Agnostic Interest Learning for Recommender Systems}

\begin{anonsuppress}
\author{Philip J. Feng}
\affiliation{%
 \institution{NetEase Cloud Music, NetEase Inc.}
 \city{Hangzhou}
 \country{China}}
\email{philipj.feng@outlook.com}

\author{Pingjun Pan}
\affiliation{%
 \institution{NetEase Cloud Music, NetEase Inc.}
 \city{Hangzhou}
 \country{China}}
\email{panpingjun@corp.netease.com}
 \authornote{Pingjun Pan contributes to model implementation and online testing.}

\author{Tingting Zhou}
\affiliation{%
 \institution{NetEase Cloud Music, NetEase Inc.}
 \city{Hangzhou}
 \country{China}}
\email{hzzhoutingting15@corp.netease.com}

\author{Hongxiang Chen}
\affiliation{%
 \institution{NetEase Cloud Music, NetEase Inc.}
 \city{Hangzhou}
 \country{China}}
\email{hzchenhongxiang@corp.netease.com}
 \authornote{Hongxiang Chen is the corresponding author.}
 
\author{Chuanjiang Luo}
\affiliation{%
 \institution{NetEase Cloud Music, NetEase Inc.}
 \city{Hangzhou}
 \country{China}}
\email{luochuanjiang03@corp.netease.com}
\end{anonsuppress}

\newcommand{\etal}{\textit{et al}. }
\newcommand{\etall}{\textit{et al}.}
\newcommand{\ie}{\textit{i}.\textit{e}.}
\newcommand{\eg}{\textit{e}.\textit{g}.}

\begin{abstract}
User behavior has been validated to be effective in revealing personalized preferences for commercial recommendations. However, few user-item interactions can be collected
for new users, which results in a nullspace for their interests, \ie, the cold-start dilemma. In this paper, a two-tower framework, namely, the model-agnostic interest learning (MAIL) framework, is proposed to address the cold-start recommendation (CSR) problem for recommender systems. In MAIL, one unique tower is constructed to tackle the CSR from a zero-shot view, and the other tower focuses on the general ranking task. Specifically, the zero-shot tower first performs cross-modal reconstruction with dual autoencoders to obtain virtual behavior data from highly aligned hidden features for new users; and the ranking tower can then output recommendations for users based on the completed data by the zero-shot tower. Practically, the ranking tower in MAIL is model-agnostic and can be implemented with any embedding-based deep models. Based on the cotraining of the two towers, the MAIL presents an end-to-end method for recommender systems that shows an incremental performance improvement. The proposed method has been successfully deployed on the live recommendation system of NetEase Cloud Music to achieve a click-through rate improvement of 13\% $\sim $ 15\% for millions of users. Offline experiments on real-world datasets also show its superior performance in CSR. Our code is available\footnote{https://github.com/LiangjunFeng/MAIL}.
\end{abstract}




\maketitle
\section{Introduction}
Recommender systems are playing important roles in various web and mobile applications, \eg, music platforms, which pick a fraction of items that a user might enjoy based on the user's past behavior and current context \cite{25,7}. Despite the success of recent deep learning \cite{30} and matrix factorization-based methods \cite{17}, a common challenge for recommender systems is the cold-start recommendation (CSR); that is, the recommendation model shows degraded performance in selecting satisfactory items for new users because of the lack of user-item interaction records \cite{9,15}.\par

Different approaches have been proposed to handle the cold-start problem. A basic CSR method is to leverage personal information to generate recommendations for new users. This is reasonable because some basic attributes of users, \eg, age and occupation, can be collected when they register accounts, and people with similar attributes usually have similar interests and behaviors \cite{2}. Some other auxiliary information is also used in CSR. For example, cross-domain knowledge is used by Hsu \etal \cite{27} and Philip \etal \cite{5} to enrich the feature representation of new users. Complementary heterogeneous information is leveraged to enrich user-item interactions by Lu \etal \cite{13} using heterogeneous information networks. A stacked denosing autoencoder is used in a hybrid model to extract embedding knowledge for CSR by Wei \etal \cite{22}. In \cite{26,10,20}, the popular meta-learning framework is applied to transfer knowledge from support sets and treat CSR as a few-shot learning task. However, the major limitation of these approaches is that new users' behaviors and interests are still missing. The auxiliary information, such as semantic knowledge and pretrained embedding, cannot change the dilemma that the recommender system knows less about new users than old users.\par
\begin{figure}[htb]
\centering
\includegraphics[width=.4\textwidth]{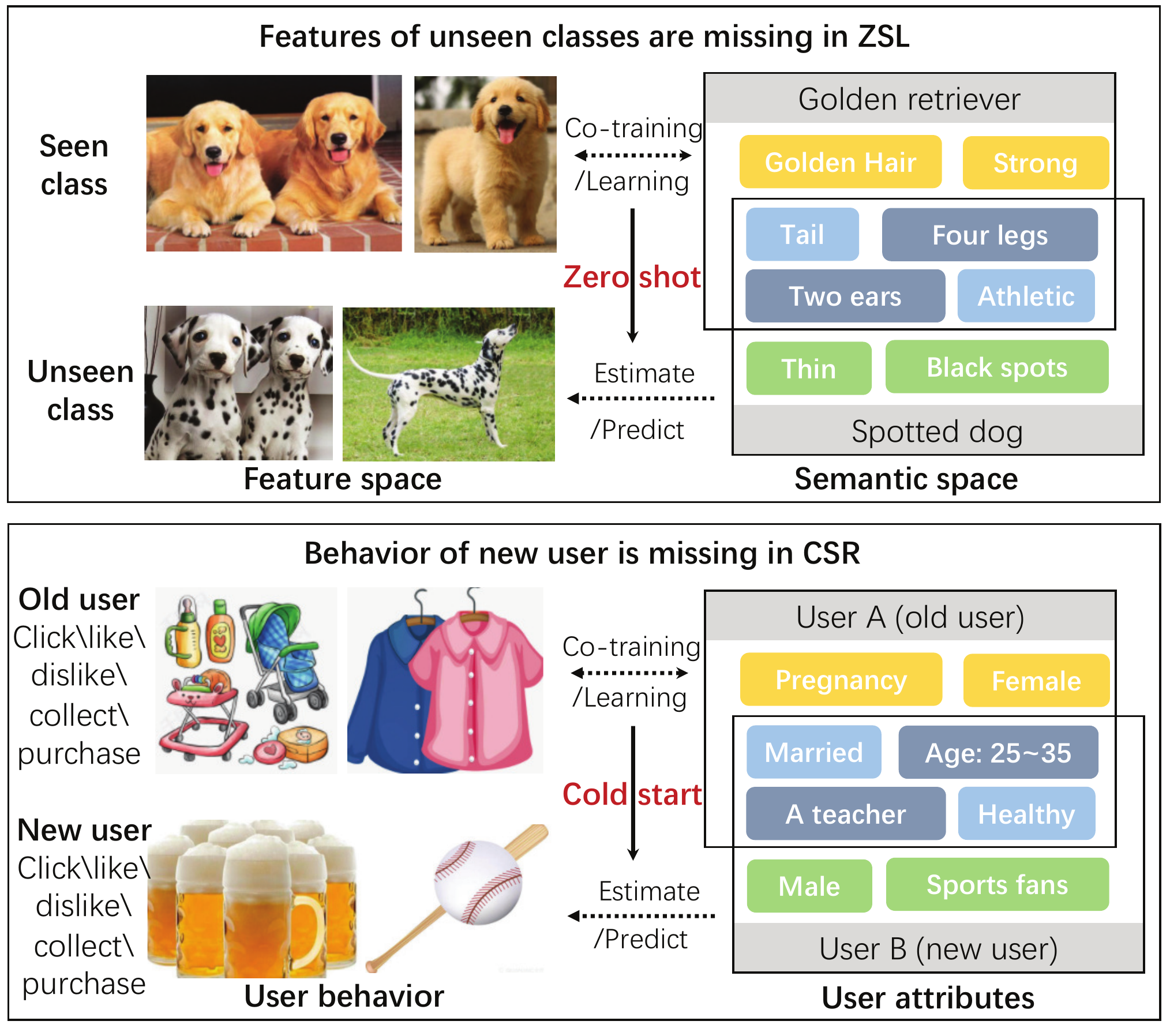} 
\caption{Paradigms of ZSL and CSR. New users' behavior in CSR is missing like features of unseen class in ZSL. Similar paradigms allow addressing CSR from a zero-shot view.}
\vspace{-2em}
\end{figure}
Recently, zero-shot learning (ZSL) \cite{6,23}, which has the same intention as CSR, has attracted ever-increasing attention in the computer vision field. Both ZSL and CSR involve the missing data problem of some objects in a specific scene. To provide a better understanding, we compare paradigms of ZSL and CSR in Figure 1. In the image recognition task, the training features of the unseen class are missing, and ZSL uses textual descriptions of objects to perform knowledge transfer from the seen class to the unseen class. In the cold-start recommendation task, behavior data of new users are missing, and CSR can utilize user attributes to generalize interests from old users to new users. Based on the zero-shot view, Li \etal \cite{4} proposed the low-rank linear autoencoder (LLAE) for CSR. The LLAE addresses CSR with a linear autoencoder, where a new user's behavior is learned from attributes with a low-rank constraint. Kumar \etal \cite{24} also applied some classic zero-shot models, \eg, ESZSL \cite{21} and ConSE \cite{14}, for unseen hashtag recommendation. However, the behavior data in CSR are different from the image features in ZSL. A fact that should be emphasized is that a user can interact with any items in the recommender system. The huge number of items in real-world applications make the modal-shift problem between user behavior and attributes more challenging than that in ZSL. The linear models discussed above may not well cover the complex mapping, and a more effective alignment technique is desired to overcome the shift problem between user behavior and attributes.\par

Moreover, we argue that a practical cold-start module is supposed to be a supplement to the existing ranking model instead of a substitute. Some methods address CSR by replacing the existing ranking model with a new method. For example, Zhao \etal \cite{19} designed a probabilistic matrix factorization method based on the similarity propagation algorithm, and Wang \etal \cite{11} used improved collaborative filtering to obtain rating information for item recommendation. The limitation of these methods is that they pay too much attention to the cold-start problem but ignore the state-of-the-art ranking performance. Although popular ranking models, \eg, ESMM \cite{16} and DMR \cite{18}, suffer from the cold-start problem on new users, they usually perform well on old users. Incremental performance improvement can be expected by integrating the cold-start module with the ranking model together.\par

Based on the above discussion, we propose a model-agnostic interest learning (MAIL) framework for recommender systems to address the cold-start problem in this paper. In MAIL, a new zero-shot tower and a general ranking tower are constructed together. From a cotraining view, the zero-shot tower formulates CSR as a ZSL task to provide virtual behavior data for new users, and the ranking tower can then capture new user's interests based on the generated virtual data. Specifically, in the zero-shot tower, two spaces, \ie, user behavior space and user attribute space, are explicitly constructed and learned by dual autoencoders. In the common hidden space of autoencoders, cross-modal reconstruction is performed to overcome the modal-shift problem between user behavior and user attributes for old users. Based on the homogeneity of user's attributes, new user's behavior can be readily reconstructed from well-aligned hidden features. For the ranking tower, it is model-agnostic and can be implemented with any embedding-based ranking models. The ranking tower is assigned to share its embedding layer with the zero-shot tower to tackle the data sparsity problem for efficient model training. We show that integrating zero-shot learning together with popular ranking models makes recommender systems free from CSR so that performance improvement on state-of-the-art results is achieved. The contributions of this paper are fourfold:\par
1) A new framework, namely, the model-agnostic interest leaning (MAIL) framework, is designed for recommender systems, which constructs a unique zero-shot tower to provide virtual behavior data for new users, enabling the ranking tower to capture their interests for cold-start recommendations.\par
2) Based on cross-modal reconstruction with dual autoencoders, the zero-shot tower in MAIL makes the hidden features of user attributes and user behavior highly aligned to generalize behavior data from old users to new users.\par
3) The ranking tower in MAIL can be implemented with any embedding-based deep models for practical application, and it also shares its embedding layer with the zero-shot tower to tackle the data sparsity problem for effective model training.\par
4) In additon to offline experiments, we deploy the proposed MAIL model on an online large-scale live recommendation system to verify the model effectiveness with millions of users in a real-world setting.\par

\section{Related Work}
\subsection{Zero-Shot Learning}
In the computer vision field, intelligent models optimize millions of parameters using sufficient training samples to obtain recognition ability for target objects \cite{1}. However, due to the long-tailed distribution of object categories, the collection of numerous samples for some rarely seen objects is generally challenging. To recognize unseen objects, zero-shot learning is explored by Lampert \etal \cite{23} and Akata \etal \cite{12} in recent years. From a macro perspective, existing ZSL methods can be grouped into three categories, including classic probability-based methods, compatibility-based methods, and generative models. Probability-based methods are represented by direct attribute-based prediction \cite{8}, as these methods learn a number of probabilistic classifiers for attributes and combine the scores to make predictions. Incremental classifiers are also used by Feng \etal \cite{33} to balance the learning between seen and unseen classes. Compatibility-based methods learn a compatibility function between the semantic embedding space and the vision feature space. For example, the ESZSL model proposed by Romera-Paredes \etal \cite{21} applies square loss and implicit regularization between embedding spaces to rank the class possibilities in an embarrassingly simple way. A hashing model is trained by Ji \etal \cite{29} to transfer knowledge and implement cross-model learning based on the semantic embedding space. Generative models train a conditional generator for visual features based on semantic features. Popular generative models include autoencoders and generative adversarial networks \cite{3}. In comparison with the other two paradigms, generative models tackle the essential insufficient data problem for ZSL to achieve state-of-the-art results. The proposed zero-shot tower in MAIL is also constructed as a generative model to tackle the missing behavior problem for CSR.
\vspace{-1em}
\subsection{Cold-Start Recommendation}
While deep ranking models have achieved considerable success in recommender systems, difficulty usually arises in contending with new users with a nullspace for their interests, known as cold-start recommendation. Some solutions in solving this problem have been proposed in current publications, including the input of semantic knowledge, second-domain knowledge transfer, active learning and meta-learning. For example, Chou \etal \cite{35} proposed a tensor factorization-based algorithm that exploits content features extracted from music audio to address CSR for the emerging application next-song recommendation. A unique characteristic of the algorithm is that it learns and updates the mapping between the audio feature space and the item latent space each time during the iterations of the factorization process, which makes the content features more effective. Li \etal \cite{2} proposed an innovative cross-domain recommendation model based on partial least squares regression. The cross-domain model is able to purely use source-domain ratings to predict the ratings for cold-start users who never rated items in the target domains. Apart from utilizing auxiliary information from other scenes and datasets, some researchers have applied active learning approaches to deal with CSR. Zhu \etal \cite{32} designed useful user selection criteria based on items' attributes and users' rating history, and combined the criteria in an optimization framework for selecting users. By exploiting the feedback ratings, accurate rating predictions for the other unselected users can then be generated. The popular meta-learning \cite{28} approach is also applied to address CSR, which regards modeling each user's preference as a learning task and minimizes the sum of the recommendation loss on training users to globally update the parameters of the meta-learner. By providing the initialization of parameters, the meta-learner guides the learning process for new users. The major difference between the proposed MAIL and the above works is that the MAIL addresses the essential missing behavior problem for new users by transferring knowledge from old users with a zero-shot tower. In addition to the basic user attributes, neither auxiliary information nor an additional dataset is added in the modeling for CSR, which is convenient when feature engineering and data collection are expensive.
\begin{figure*}[htb]
\centering
\includegraphics[width=0.9\textwidth]{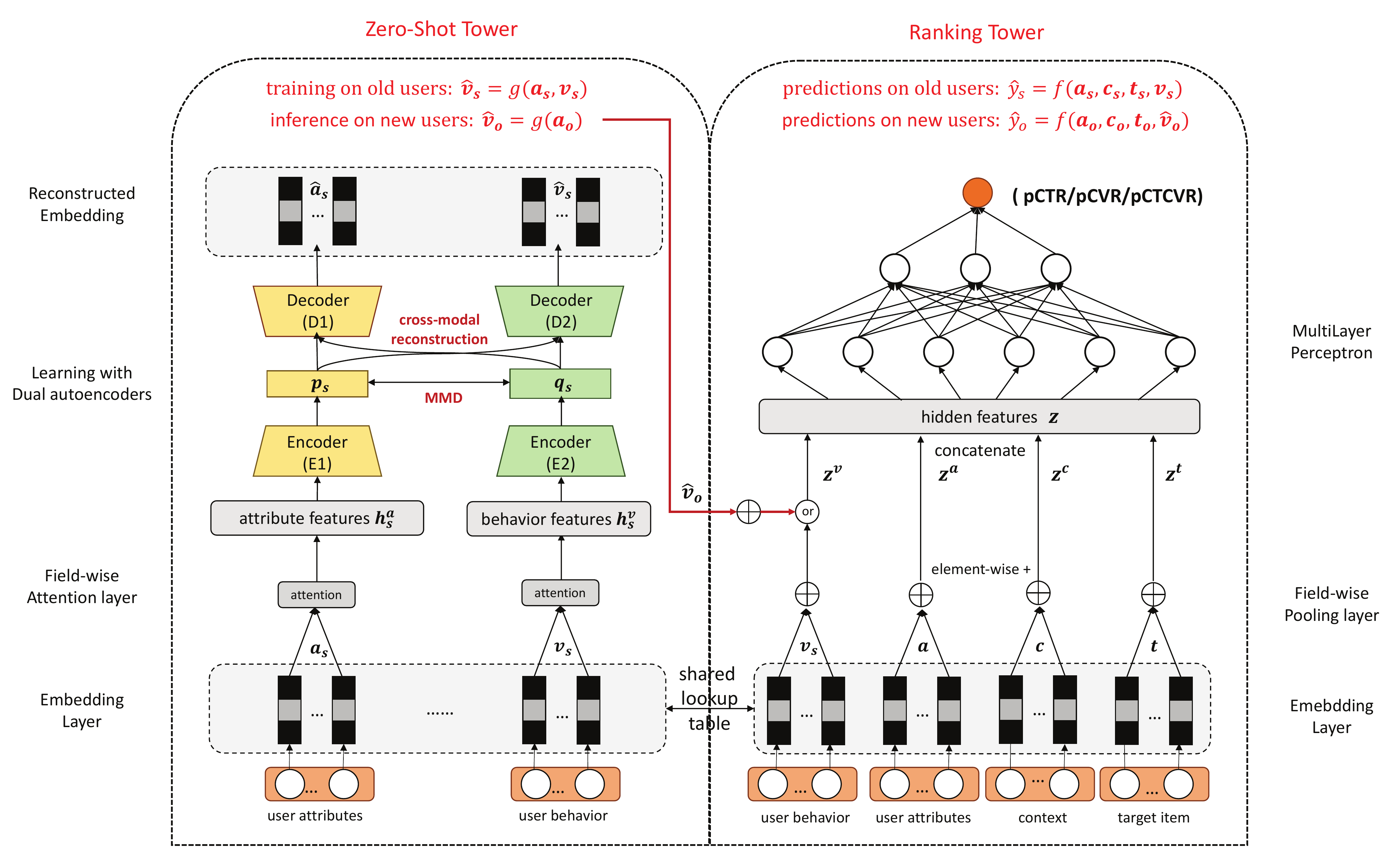} 
\caption{Illustration of the proposed MAIL, where a two-tower structure can be observed. The zero-shot tower addresses the CSR by providing virtual new user's behavior data $\hat{\bm{v}}_{o}$ for the ranking tower. The ranking tower outputs recommendations based on the completed data and shares its embedding lookup table with the zero-shot tower.}
\vspace{-1em}
\end{figure*}
\section{Proposed Approach}
\subsection{Problem Formulation and Notations}
\subsubsection{Features in Recommender Systems}
There are usually four categories of features in recommender systems: user attributes $\bm{a} \in \mathcal{A}$, user behavior $\bm{v} \in \mathcal{V}$, context $\bm{c} \in \mathcal{C}$, and target item $\bm{t} \in \mathcal{T}$. User attributes in $\mathcal{A}$ contain the user ID, occupation, and so on; user behavior in $\mathcal{V}$ is a sequential list of user interacted items with corresponding IDs. Context in $\mathcal{C}$ contains the date, recall method, location, etc. Features of target item in $\mathcal{T}$ are the item ID, category ID, and so on. The ranking model in a recommender system learns a mapping function $f(\bm{x})$ to output the score $\hat{y}$ for an item, where $\bm{x} = \{\bm{a},\bm{v},\bm{c},\bm{t}\}$ is the input. Generally, items with higher scores will be recommended to users with priority.\par
Note that most of the mentioned features are categorical, which are represented by one-hot vectors with high dimension. In deep learning models, one-hot features are usually transformed into low-dimensional dense features by an embedding layer. Hence, the notations, \eg, $\bm{a}$ and $\bm{v}$, actually denote the embedding vectors. For example, user attributes are $\bm{a} = [\bm{a}_{1},...,\bm{a}_{n_{a}}] \in \mathbb{R}^{n_{a} \times d}$, where $n_{a}$ is the number of attributes and $d$ is the embedding dimension. For the user behavior sequence $\bm{v} = [\bm{v}_{1},...,\bm{v}_{n_{v}}] \in \mathbb{R}^{n_{v} \times d}$, each item $\bm{v}_{i}$ is obtained by performing element-wise multiplication between the embedding of the item ID and the embedding of the user's action. Moreover, some dense features can be directly used without the embedding transformation.
\vspace{-0.5em}
\subsubsection{Zero-Shot Learning for CSR}
In cold-start recommendation, we have two scopes: the old user scope $\mathcal{S}$ and the new user scope $\mathcal{O}$. For old users, four categories of features are ready for ranking models, which are denoted as $\bm{x}_{s} = \{\bm{a}_{s},\bm{v}_{s},\bm{c}_{s},\bm{t}_{s}\}$. For new users, the behavior $\bm{v}_{o}$ is missing, and features of new users are denoted as $\bm{x}_{o} = \{\bm{a}_{o},\bm{c}_{o},\bm{t}_{o}\}$. In this paper, we train a zero-shot tower $\hat{\bm{v}}_{s} = g(\bm{a}_{s}, \bm{v}_{s})$ based on old users, which can be applied to new users to obtain virtual behavior data $\hat{\bm{v}}_{o} = g(\bm{a}_{o})$ to tackle the cold-start problem. The notations $\hat{\bm{v}}_{s}=g(\bm{a}_{s}, \bm{v}_{s})$ and $\hat{\bm{v}}_{o} = g(\bm{a}_{o})$ denote that only old users' behavior data $\bm{v}_{s}$ are available for the zero-shot tower while new users' behavior data are not available. To summarize, the whole model in this paper can be formulated as $\hat{y}_{s}=F(\bm{x}_{s}) = f(\bm{a}_{s},\bm{c}_{s},\bm{t}_{s}, \bm{v}_{s})$ for old users and $\hat{y}_{o}=F(\bm{x}_{o}) = f(\bm{a}_{o},\bm{c}_{o},\bm{t}_{o}, \hat{\bm{v}}_{o})$ for new users, where $\hat{\bm{v}}_{o} = g(\bm{a}_{o})$.
\subsection{Model-Agnostic Zero-Shot Interest Learning}
\subsubsection{Overall Idea} The overall idea of the proposed MAIL is illustrated in Figure 2. In addition to a general ranking tower, a new zero-shot tower can be observed. The ranking tower outputs recommendations based on the completed data by the zero-shot tower, and the zero-shot tower enjoys the dense embedding by the ranking tower. The cotraining of the two towers make the recommender systems free from CSR so that an incremental performance improvement is achieved. 
\subsubsection{Zero-Shot Tower for Cold-Start Problem}
Here, we show how to construct the zero-shot tower for CSR in detail.\par
First, it is clarified that the zero-shot tower is trained on old users and makes inference on new users to transfer behavior data from old users to new users based on their attributes. Hence, two kinds of embedding features, \ie, user attributes $\bm{a}_{s}$ and user behavior $\bm{v}_{s}$, are used for model training. The embedding features are trained by the ranking tower. Through learning from the dense embedding vectors, the zero-shot tower obtains more effective training and extraction processes.\par
To weight different features, the self-attention technique is then applied on $\bm{a}_{s}$ and $\bm{v}_{s}$ to construct attribute features $\bm{h}^{a}_{s}$ and behavior features $\bm{h}^{v}_{s}$, respectively. Taking $\bm{h}^{a}_{s}$ as an example, the attention mechanism is performed as follows:
\begin{equation}
\begin{aligned}
\bm{h}^{a}_{s} = \sum^{n_{a}}_{i=1}\alpha_{i}\bm{a}_{si},
\end{aligned}
\end{equation}
where $n_{a}$ is the number of user attributes and $\bm{a}_{si} \in \mathbb{R}^{d}$ is the embedding vector of the $i$-th user attribute in $\bm{a}_{s}$. The attention weight $\alpha_{i}$ is obtained by
\begin{equation}
\begin{aligned}
\alpha_{i} &= \frac{exp(e_{i})}{\sum^{n_{a}}_{j=1}e_{j}}, \\
e_{i} =\;& \bm{z}^{T}tanh(\bm{w}\bm{a}_{si}+\bm{b}),
\end{aligned}
\end{equation}
where $\bm{z} \in \mathbb{R}^{d_{r}}$, $\bm{w} \in \mathbb{R}^{d_{r} \times d}$, and $\bm{b} \in \mathbb{R}^{d_{r}}$ are trainable parameters, and $d_{r}$ is the given attention dimension.\par
Based on $\bm{h}^{a}_{s}$ and $\bm{h}^{v}_{s}$, cross-modal reconstruction is performed to align user attributes and behavior for a more effective attribute-based behavior generation. In the attribute space, the cross-modal reconstruction loss is:
\begin{equation}
\begin{aligned}
\mathcal{L}_{a} = ||D1(\bm{p}_{s})-\bm{a}_{s}||^{2}_{2} + ||D2(\bm{p}_{s})-\bm{v}_{s}||^{2}_{2},
\end{aligned}
\end{equation}
and in the behavior space, the cross-modal reconstruction loss is:
\begin{equation}
\begin{aligned}
\mathcal{L}_{v} = ||D1(\bm{q}_{s})-\bm{a}_{s}||^{2}_{2} + ||D2(\bm{q}_{s})-\bm{v}_{s}||^{2}_{2},
\end{aligned}
\end{equation}
where $\bm{p}_{s} = E1(\bm{h}^{a}_{s}) \in \mathbb{R}^{d_{h}}$, $\bm{q}_{s} = E2(\bm{h}^{v}_{s}) \in \mathbb{R}^{d_{h}}$, and $d_{h}$ is the given hidden dimension. $E1$ and $E2$ are encoders for user attributes and user behavior, and $D1$ and $D2$ are corresponding decoders, which are shown in Figure 2.\par

The cross-modal reconstruction loss assigns the hidden features $\bm{p}_{s}$ and $\bm{q}_{s}$ to reconstruct $\bm{a}_{s}$ and $\bm{v}_{s}$, respectively, with the common decoders, which actually makes $\bm{p}_{s}$ and $\bm{q}_{s}$ have the same prediction ability for user attributes and user behavior. Take the exposition a step further, we denote $\bm{P}_{s} \in \mathbb{R}^{n \times d_{h}}$ as a batch of $\bm{p}_{s}$,  $\bm{Q}_{s} \in \mathbb{R}^{n \times d_{h}}$ as a batch of $\bm{q}_{s}$, and $n$ as the batch size. The maximum mean discrepancy (MMD) constraint is then applied to $\bm{P}_{s}$ and $\bm{Q}_{s}$ to minimize their distribution distance and guide the convergence of cross-modal reconstruction, which can be formulated as:
\begin{equation}
\begin{aligned}
\mathcal{L}_{d} &= D_{\mathcal{H}}(\bm{P}_{s},\bm{Q}_{s}) = \left \| \frac{1}{n}\sum^{n}_{i=1}\phi(\bm{p}_{si}) - \frac{1}{n}\sum^{n}_{j=1}\phi(\bm{q}_{sj})   \right \|_{2}^{2} \\
                &= \frac{1}{n}\sum^{n}_{i=1}\sum^{n}_{j=1}k(\bm{p}_{si},\bm{p}_{sj}) + \frac{1}{n}\sum^{n}_{i=1}\sum^{n}_{j=1}k(\bm{q}_{si},\bm{q}_{sj}) \\
                & - \frac{2}{n^{2}}\sum^{n}_{i=1}\sum^{n}_{j=1}k(\bm{p}_{si},\bm{q}_{sj}),
\end{aligned}
\end{equation}
where $\mathcal{H}$ is the reproducing kernel Hilbert space (RKHS), $\phi:\bm{p}_{s}, \bm{q}_{s} \rightarrow \mathcal{H}$, $k$ is a Gaussian kernel function:
\begin{equation}
\begin{aligned}
k(\bm{p}_{s}, \bm{q}_{s}) = exp(-\left \| \bm{p}_{s}-\bm{q}_{s} \right \|^{2}/2\sigma^{2}),
\end{aligned}
\end{equation}
and $\sigma=1$ is the bandwidth in this paper.\par
To summarize, the training loss of the zero-shot tower is:
\begin{equation}
\begin{aligned}
\mathcal{L}_{zst} = \mathcal{L}_{a} + \mathcal{L}_{v} + \mathcal{L}_{d}.
\end{aligned}
\end{equation}
The first two items in $\mathcal{L}_{zst}$ require both $\bm{p}_{s}$ and $\bm{q}_{s}$ to have the reconstruction ability for $\bm{a}_{s}$ and $\bm{v}_{s}$, while the third item in $\mathcal{L}_{zst}$ makes $\bm{p}_{s}$ and $\bm{q}_{s}$ enjoy the same hidden space for a better convergence of cross-modal reconstruction. With the triple-aligned hidden features, the virtual behavior data can be readily generated for new users from their attributes as follows:
\begin{equation}
\begin{aligned}
\hat{\bm{v}_{o}} = D2(\bm{p}_{o}) = D2(E1(\bm{h}_{o}^{a})), 
\end{aligned}
\end{equation}
where $\bm{h}_{o}^{a}$ is the attention weighted features of $\bm{a}_{o}$ and can be obtained by Eqs. (1) and (2).
\subsubsection{Ranking Tower for Recommendation}
Here, we show how to construct the ranking tower for recommendation in detail.\par
Generally, the ranking tower is model-agnostic and can be implemented by any embedding-based models for estimating the click-through rate (CTR), post-click conversion rate (CVR), or post-view click through\&conversion rate (CTCVR) \cite{16}. Here, the basic deep model shown in Figure 2 is introduced to provide a better understanding.\par
The ranking tower treats new users and old users in different ways, which means there should be a \emph{flag} to indicate whether a user is a new user, which can be obtained by
\begin{equation}
\begin{aligned}
\emph{flag} = \left\{\begin{matrix}
true  & \bm{v} = \varnothing  \\ 
false & otherwise,
\end{matrix}\right.
\end{aligned}
\end{equation}
where $\emph{flag} = true$ for new users and $\emph{flag} = false$ for old users.\par
Based on the embedding lookup table and \emph{flag}, the input $\bm{x}$ of the ranking tower can be denoted as: 
\begin{equation}
\begin{aligned}
\bm{x} = \left\{\begin{matrix}
\{\bm{a}_{o},\bm{c}_{o},\bm{t}_{o},\hat{\bm{v}}_{o}\}  & \emph{flag} = true  \\ 
\{\bm{a}_{s},\bm{c}_{s},\bm{t}_{s},\bm{v}_{s}\}        &\; \emph{flag} = false,
\end{matrix}\right.  
\end{aligned}
\end{equation}
where $\hat{\bm{v}}_{o}$ obtained from Eq. (8) is the virtual behavior data for new users.\par
A basic sum pooling layer is then added to obtain the concatenated hidden features $\bm{z} = [\bm{z}^{v}, \bm{z}^{a}, \bm{z}^{c}, \bm{z}^{t}] \in \mathbb{R}^{4d}$ from $\bm{x}$, where $d$ is the embedding dimension. Taking $\bm{z}^{a}$ as an example, the sum pooling is performed as follows:
\begin{equation}
\begin{aligned}
\bm{z}^{a} = \sum^{n_{a}}_{i=1} \bm{a}_{i},
\end{aligned}
\end{equation}
where $\bm{a}_{i}$ is the $i$-th attribute embedding vector in $\bm{a}$.\par
The final output $\hat{y}$ is obtained by a multilayer perceptron, which is formulated as:
\begin{equation}
\begin{aligned}
\bm{h}_{z1} =  Lea&kyReLU(BN(Dense(Dropout(\bm{z},0.5)))) \\
\bm{h}_{z2} =  Leak&yReLU(BN(Dense(Dropout(\bm{h}_{z1},0.5)))) \\
& \hat{y} =     Sigmoid(Dense(\bm{h}_{z2}))), 
\end{aligned}
\end{equation}
and the training loss of the ranking tower is:
\begin{equation}
\begin{aligned}
\mathcal{L}_{rt} = -\sum^{n}_{i=1}(y_{i}log\hat{y}_{i} + (1-y_{i})log(1-\hat{y}_{i})),
\end{aligned}
\end{equation}
where $n$ is the batch size.\par
In summary, the overall training loss of MAIL is:
\begin{equation}
\begin{aligned}
\mathcal{L} = \mathcal{L}_{zst} + \mathcal{L}_{rt},
\end{aligned}
\end{equation}
where two towers are cotrained to tackle the CSR for recommender systems. It is worth mentioning that, in practical applications, not only will there be new users in the test data but there will also be new users in the training data. It is difficult for a ranking and CSR two-stage model to address the missing data problem in the ranking model's training phase because the CSR model based on embedding features has to be trained after the ranking model. In contrast, the proposed MAIL is designed as an end-to-end model to obtain virtual behavior data from the zero-shot tower for the ranking tower whenever the need arises and is hence more practical for applications.
\begin{figure}[htb]
\centering
\includegraphics[width=0.4\textwidth]{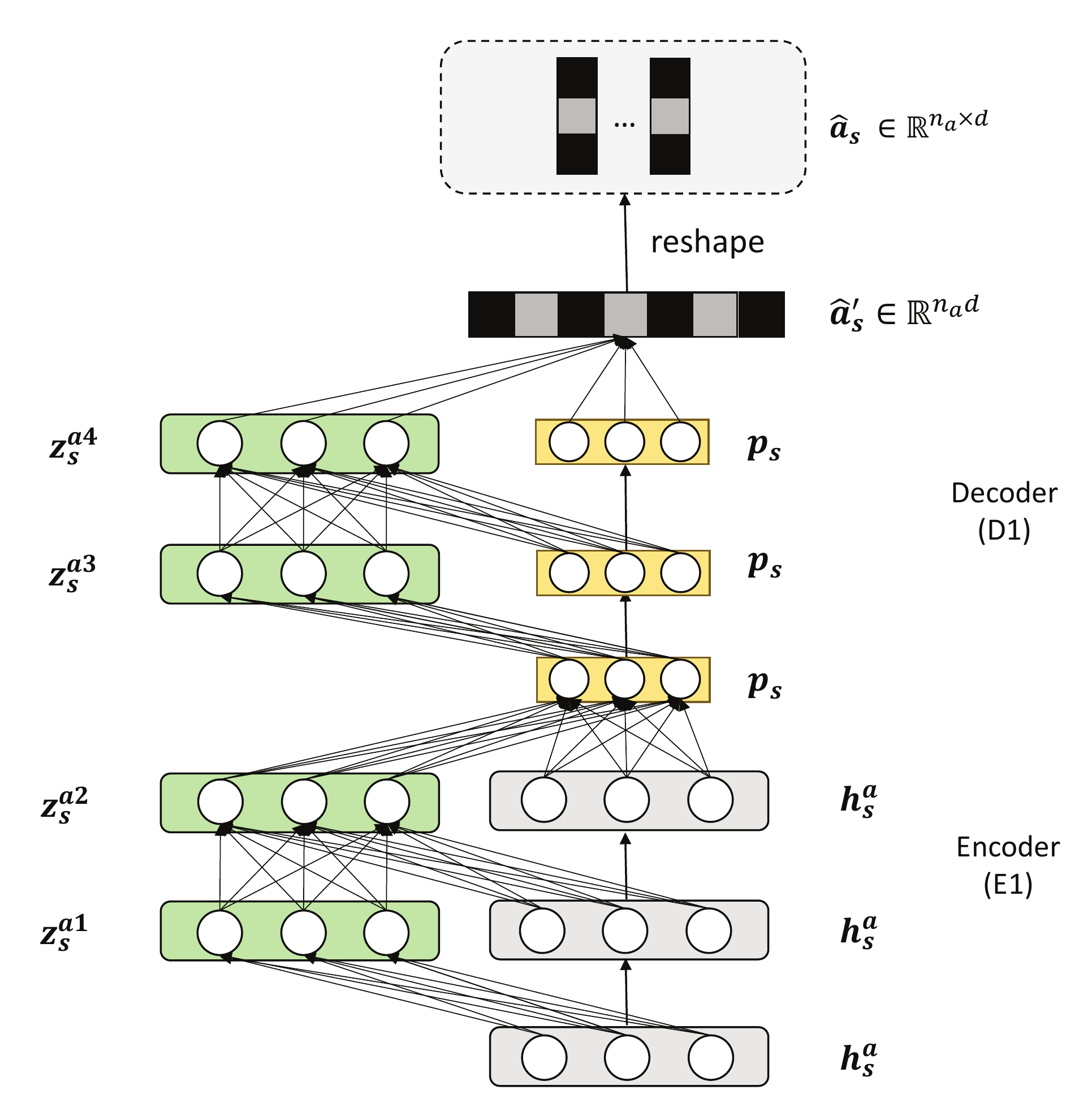} 
\caption{Illustration of the autoencoder used in MAIL, where the encoder and decoder are designed with residual structures for efficient feature extraction.}
\vspace{-1em}
\end{figure}
\subsubsection{Implementation Tricks and the Algorithm of MAIL}
Here, implementation tricks and the training algorithm of MAIL are summarized.\par
1) Cotraining with two optimizers: The two towers in MAIL are trained by two optimizers, respectively. One optimizer $G_{A}$ is used for $\mathcal{L}_{zst}$, and the other  $G_{B}$ is for $ \mathcal{L}_{rt}$. It is noted that the embedding lookup table is completely trained by $G_{B}$ together with the ranking tower and is not affected by $G_{A}$. Because the aim of the zero-shot tower is to reconstruct embedding vectors instead of changing them, applying the gradients of the zero-shot tower on the embedding lookup table would degrade model performance.\par
2) Autoencoders with residual structure: Two autoencoders are used in the zero-shot tower to perform the cross-modal reconstruction. Considering the sparse information in recommender systems, a residual structure is designed for the autoencoder to perform the reconstruction task better. Taking the left autoencoder in Figure 2 as an example, the residual structure is shown in Figure 3. The forward-propagation of the residual encoder can be formulated as follows:
\begin{equation}
\begin{aligned}
&\bm{z}^{a1}_{s} =  LeakyReLU(Dense(Dropout(\bm{h}^{a}_{s},0.5))) \\
\bm{z}^{a2}_{s}  =  &LeakyReLU(Dense(Dropout(Concat(\bm{z}^{a1}_{s},\bm{h}^{a}_{s}),0.5))) \\
& \bm{p}_{s} =     LeakyReLU(Dense(Concat(\bm{z}^{a2}_{s},\bm{h}^{a}_{s})))), 
\end{aligned}
\end{equation}
and the forward-propagation of the residual decoder can be formulated as:
\begin{equation}
\begin{aligned}
\bm{z}^{a3}_{s}& =  LeakyReLU(Dense(Dropout(\bm{p}_{s},0.5))) \\
\bm{z}^{a4}_{s}  =  Leaky&ReLU(Dense(Dropout(Concat(\bm{z}^{a3}_{s},\bm{p}_{s}),0.5))) \\
& {\hat{\bm{a}}}'_{s} =     Dense(Concat(\bm{z}^{a4}_{s},\bm{p}_{s})).
\end{aligned}
\end{equation}
The principle of the residual structure is similar to that of the wide\&deep model \cite{34}, where both the wide module's memory ability and the deep module's generalization ability are retained by feature concatenation.\par
3) Training algorithm of MAIL: The algorithm of MAIL is summarized in Algorithm 1.
\begin{table}[!htb]
\centering
    \setlength{\tabcolsep}{2 mm}
    \centering
    \begin{tabular}{ll}
        \Xhline{1pt}
        \multicolumn{2}{l}{\textbf{Algorithm 1} Learning Procedure for MAIL}\\
        \hline
        \multicolumn{2}{l}{\textbf{Zero-Shot Tower, }}\\
        \multicolumn{2}{l}{\tabincell{l}{\qquad Input -> Output for Old Users: \qquad \, $g(\bm{a}_{s},\bm{v}_{s}) = \hat{\bm{v}}_{s}$ }} \\
        \multicolumn{2}{l}{\tabincell{l}{\qquad Input -> Output for New Users: \qquad $g(\bm{a}_{o}) = \hat{\bm{v}}_{o}$ }} \\
        \multicolumn{2}{l}{\textbf{Ranking Tower,}}\\
        \multicolumn{2}{l}{\tabincell{l}{\qquad Input -> Output for Old Users: \quad \;\;\;\,$f(\bm{a}_{s}, \bm{c}_{s}, \bm{t}_{s}, \bm{v}_{s}) = \hat{y}_{s}$ }} \\
        \multicolumn{2}{l}{\tabincell{l}{\qquad Input -> Output for New Users: \quad \;\;\,$f(\bm{a}_{o}, \bm{c}_{o}, \bm{t}_{o}, \hat{\bm{v}}_{o}) = \hat{y}_{o}$ }} \\
        \hline
        \multicolumn{2}{l}{\textbf{Training Phase:}}\\
        1    & \tabincell{l}{Construct the computing graph as Figure 2 and initialize\\ the model parameters using the Xavier method.}  \\
        2    & \tabincell{l}{Two Adam optimizers are used for model optimization. \\ Optimizer $G_{A} \rightarrow \mathcal{L}_{zst}$; Optimizer $ G_{B} \rightarrow \mathcal{L}_{rt}$.}  \\
        \multicolumn{2}{l}{\;\;\;\;\;\;\; \textbf{For $i=1:I$ (number of epochs) do}}\\
        \multicolumn{2}{l}{\;\;\;\;\;\;\;\;\;\;\; \textbf{For $j=1:J$ (rounds in an epoch) do}}\\ 
        3   & \tabincell{l}{\;\;\;\;\;\; Make \emph{flag} for a batch of data by Eq. (9).}\\ 
        \multicolumn{2}{l}{\;\;\;\;\;\;\;\;\;\;\;\;\;\; \textbf{Zero-Shot Tower:}}\\
        4   & \tabincell{l}{\;\;\;\;\;\; Obtain $\mathcal{L}_{zsl}$ by Eq. (7) for users with $\emph{flag} = false$.}\\ 
        5   & \tabincell{l}{\;\;\;\;\;\; Perform back-propagation and apply $G_{A}$ for $\mathcal{L}_{zsl}$.}\\ 
        6   & \tabincell{l}{\;\;\;\;\;\; Obtain $\hat{\bm{v}}_{o}$ by Eq. (8) for users with $\emph{flag} = true$.}\\ 
        \multicolumn{2}{l}{\;\;\;\;\;\;\;\;\;\;\;\;\;\; \textbf{Ranking Tower:}}\\
        7   & \tabincell{l}{\;\;\;\;\;\; If $\emph{flag}$, $\bm{x} = \{\bm{a}_{o}, \bm{c}_{o}, \bm{t}_{o}, \hat{\bm{v}}_{o}\}$; else $\bm{x} = \{\bm{a}_{s}, \bm{c}_{s}, \bm{t}_{s}, \bm{v}_{s}\}$}\\   
        8   & \tabincell{l}{\;\;\;\;\;\; Obtain $\mathcal{L}_{rt}$ by Eq. (13) for all users.}\\
        9   & \tabincell{l}{\;\;\;\;\;\; Perform back-propagation and apply $G_{B}$ for $\mathcal{L}_{rt}$.}\\
            & \tabincell{l}{\;\;\; \textbf{End for}}\\
        \multicolumn{2}{l}{\;\;\;\;\;\;\; \textbf{End for}}\\
        \hline
        \multicolumn{2}{l}{\textbf{Testing Phase:}}\\
        10   & \tabincell{l}{Make \emph{flag} for all users by Eq. (9).}\\
        \multicolumn{2}{l}{\;\;\;\;\;\;\; \textbf{Zero-Shot Tower:}}\\
        11   & \tabincell{l}{Obtain $\hat{\bm{v}}_{o}$ for users with $\emph{flag}=true$ by Eq. (8)}.\\
        \multicolumn{2}{l}{\;\;\;\;\;\;\; \textbf{Ranking Tower:}}\\
        12   & \tabincell{l}{Obtain $\hat{y}$ for all users by Eq. (12).}\\
        \Xhline{1pt}  
    \end{tabular}
\vspace{-0.5em}
\end{table}
\subsection{Feasibility Analysis of Attribute-based Knowledge Transfer}
In MAIL, the zero-shot tower generalizes behavior data from old users to new users based on their attributes. A basic assumption behind the knowledge transfer is that people with similar attributes have similar behaviors. However, there seems to be a contradiction in that the user attribute space is much smaller than the user behavior space in recommender systems. For better clarification, we denote the number of attributes as $n_{a}$ and assume that each attribute has $k_{a}$ different values. Similarly, the length of a user's behavior sequence is denoted as $n_{v}$ and the number of items that can be interacted with users is denoted as $k_{v}$. The condition $n_{a}k_{a} >= n_{v}k_{v}$ seems to be required for effective learning from attributes to behavior, while both $n_{v}$ and $k_{v}$ may be larger than $n_{a}$ and $k_{a}$, respectively, in a real recommender system.\par
Here, we highlight that the core of user behavior is user interest instead of items. Compared with learning what items users have interacted with, it is better to learn what style of items users like for recommender systems. This is one of the important reasons why the dense embedding vector $\bm{v}$ is implemented as the learning target of the zero-shot tower. Based on the dense vector, a user's interest $\phi(\bm{r})$ can be readily represented by a set of items as 
\begin{equation}
\begin{aligned}
\phi(\bm{r}) = \{\bm{r} \;|\; ||\bm{r} - \bm{v}||^{2}_{2} <=\epsilon \},
\end{aligned}
\end{equation}
where $\epsilon$ is an error bound representing the interest range. As long as the generated $\hat{\bm{a}} \in \phi(\bm{r})$, it is feasible for the ranking tower to capture user interest. Hence, the object of the user attribute space is actually the user interest space instead of the user behavior space. Meanwhile, the kinds of user interests $n_{r}$ are much smaller than $n_{v}k_{v}$, which makes the attribute-based knowledge transfer available for CSR.
\begin{table}[!htb]
\vspace{-1.5em}
\centering
    \textbf{\caption{Statistics of the Two Datasets}}
    \vspace{-1em}
    \setlength{\tabcolsep}{2 mm}
    \centering
    \begin{tabular}{cccccccc}
        \Xhline{1pt}
        Dataset  &  \textbf{Public} & \textbf{Industrial}  \\
        \Xhline{1pt}
        \# All users &  1,141,729 & 613,221 \\
        \# New users &  456,691  & 282,081 \\
        \# Items &  846,811  & 52,831  \\
        \# User attributes  & 9 & 21 \\
        \# Context features & 4 & 75 \\
        \# Avg. length of behavior sequence & 63 & 79\\
        \# Item features    & 6  & 105 \\
        \Xhline{1pt}
        \end{tabular}
\vspace{-1em}
\end{table}
\begin{table*}[thb]
\centering
    \textbf{\caption{Results on Public and Industrial Datasets}}
    \vspace{-0.5em}
    \setlength{\tabcolsep}{0.8 mm}
    \centering
    \begin{tabular}{c|cc|cc|cc|cc|cc|cc|cc|cc}    
        \Xhline{1pt}
        \multirow{3}*{\textbf{Method}} & \multicolumn{8}{c|}{\textbf{Public Dataset}} & \multicolumn{8}{c}{\textbf{Industrial Dataset}} \\
                                 \cline{2-17}
                                & \multicolumn{4}{c|}{New Users} & \multicolumn{4}{c|}{Old Users} & \multicolumn{4}{c|}{New Users} & \multicolumn{4}{c}{Old Users} \\
                                \cline{2-17}
                                & AUC & RI    & GAUC & RI   & AUC  & RI   & GAUC & RI   & AUC  & RI  & GAUC & RI   & AUC  & RI  & GAUC & RI \\
        \hline
        EmbLR           & 0.5800  & 0.00\%   & 0.5607  & 0.00\% & 0.6252  & 0.00\%    & 0.5892  & 0.00\%  & 0.6682  & 0.00\%   & 0.6365 & 0.00\%    & 0.7728 & 0.00\%   & 0.6419 & 0.00\%   \\
        LLAE            & 0.5854  & 0.93\%   & 0.5646  & 0.69\% & 0.6253  & {\color{blue}{0.01\%}}    & 0.5892  & 0.00\%  & 0.6713  & 0.46\%   & 0.6384 & 0.29\%    & 0.7737 & {\color{red}{0.11\%}}  & 0.6422 & {\color{blue}{0.05\%}}   \\
        \hline
        BaseDNN         & 0.5862   & 0.00\%   & 0.5658  & 0.00\% & 0.6271  & 0.00\%   & 0.5897  & 0.00\%  & 0.6771  & 0.00\%   & 0.6380 & 0.00\%    & 0.7782 & 0.00\%   & 0.6459 & 0.00\%    \\
        MetaEmb         & 0.5911   & 0.84\%   & 0.5682  & 0.42\% & {\color{blue}{0.6273}}  & {\color{red}{0.03\%}}   & 0.5900 & {\color{blue}{0.05\%}}  & 0.6834  & 0.93\%   & 0.6404 & 0.37\%    & 0.7788 & {\color{blue}{0.07\%}}   & 0.6466 & {\color{red}{0.10\%}}      \\
        MAIL-Base       & {\color{blue}{0.5934}}   & {\color{blue}{1.22\%}}   & {\color{blue}{0.5710}}  & {\color{blue}{0.91\%}} & 0.6271  & 0.00\%    & 0.5898  & 0.02\%  & {\color{blue}{0.6849}}  & {\color{red}{1.15\%}}   & 0.6413 & {\color{blue}{0.51\%}}   & 0.7788 & {\color{blue}{0.07\%}}   & 0.6460 & 0.01\%   \\
        \hline
        DMR             & 0.5879   & 0.00\%  & 0.5668  & 0.00\% & {\color{red}{0.6302}}  & 0.00\%   & {\color{blue}{0.5911}}  & 0.00\%  & 0.6823  & 0.00\%   & {\color{blue}{0.6420}} & 0.00\%    & {\color{blue}{0.7833}} & 0.00\%   & {\color{blue}{0.6551}} & 0.00\%      \\
        MAIL-DMR        & {\color{red}{0.5958}}   & {\color{red}{1.34}}\%  & {\color{red}{0.5733}}  & {\color{red}{1.14\%}} & {\color{red}{0.6302}}  & 0.00\%   & {\color{red}{0.5915}}  & {\color{red}{0.06\%}} & {\color{red}{0.6895}}  & 1.05\%   & {\color{red}{0.6465}} & {\color{red}{0.70\%}}    & {\color{red}{0.7838}} & 0.06\%   & {\color{red}{0.6558}} & {\color{red}{0.10\%}}   \\
        \Xhline{1pt}
        \multicolumn{17}{l}{ \tabincell{l}{{\color{red}{Red font}} and {\color{blue}{blue font}} are the highest and the second highest results in each column. EmbLR is the baseline of LLAE; BaseDNN\\ is the baseline of MetaEmb and MAIL-Base; DMR is the baseline of MAIL-DMR.}}
        \end{tabular}
        \vspace{-1em}
\end{table*}
\section{Experiments}
Here, the MAIL is validated on both public and industrial datasets. First, we introduce experimental settings and compared methods in detail. Next, the recommendation performance and an ablation study are given to show the state-of-the-art results. Visualized features are also presented to provide an intuitive understanding.
\subsection{Experimental Settings and Compared methods}
\subsubsection{Datasets} Two large-scale real-world dataset are used here. \par
1) \textbf{Public Dataset} The Alimama Dataset\footnote{\href{https://tianchi.aliyun.com/dataset/dataDetail?dataId=56}{https://tianchi.aliyun.com/dataset/dataDetail?dataId=56}} contains ad displays and click logs randomly sampled from Taobao over 8 days. It includes 1.14 million users and 0.84 million items with 6 million user-item interaction logs. The logs from the first seven days are used as the training set, and logs from the last day are used as the testing set. Note that the dataset is provided without new users. Hence, we randomly select forty percent of users in the training and testing set and treat them as new users by removing their behaviors. \par
2) \textbf{Industrial Dataset} The industrial dataset from NetEase Cloud Music contains live recommendation and user behavior logs randomly sampled from a company (abiding by the double-blind principle) over 8 days. It includes 613 thousand users and 52 thousand items with 2 million user-item interaction logs. In the industrial dataset, approximately half of the users are new users. The statistics of the two datasets are summarized in Table \uppercase\expandafter{\romannumeral1}.
\subsubsection{Compared Methods} Seven models are compared.\par
1) \textbf{EmbLR} Logistic regression is a classic linear model that can be regarded as a shallow neural network. Here, we implement logistic regression with an embedding layer, \ie, EmbLR, to content with sparse IDs. EmbLR is the baseline of LLAE.\par
2) \textbf{LLAE (2019)} LLAE \cite{4} is a zero-shot learning-based cold-start method, that generates behavior data for logistic regression with a linear low-rank autoencoder. As discussed in \cite{4}, the weight of low-rank loss is decided by cross-validation and set to 30.\par
3) \textbf{BaseDNN} BaseDNN is the ranking tower shown in Figure 2, which is used here as a baseline of MetaEmb and MAIL-Base.\par
4) \textbf{MetaEmb (2019)} MetaEmb \cite{28} is a meta-learning-based cold-start method, that learns to generate desirable initial embeddings for new IDs. The base model is implemented as the BaseDNN.\par
5) \textbf{MAIL-Base (Ours)} MAIL-Base is the proposed two-tower model shown in Figure 2, which uses the zero-shot tower to tackle the cold-start problem for BaseDNN.\par
6) \textbf{DMR (2020)} DMR \cite{18} is a state-of-the-art model for recommender systems, that consists of an item-to-item network and a user-to-item network to combine the thought of collaborative filtering for the ranking task. DMR is the baseline of MAIL-DMR.\par
7) \textbf{MAIL-DMR (Ours)} MAIL is the proposed two-tower model, that uses DMR as the ranking tower to obtain an incremental performance improvement based on the designed zero-shot tower.\par
In regard to the above seven models, EmbLR and LLAE are linear models, BaseDNN, MetaEmb, and MAIL-Base are basic deep models, and DMR and MAIL-DMR are improved deep models.
\subsubsection{Metrics} The area under the ROC curve (AUC) and the group area under the ROC curve (GAUC) are used as metrics. The model performance on both new users and old users are reported. In addition, for each result, the relative improvement (RI) between the baseline and the improved method is provided for an intuitive comparison. RI is obtained as follows:\par
\begin{equation}
\begin{aligned}
RI(a,b) = \frac{b-a}{a}*100\%,
\end{aligned}
\end{equation}
where $a$ is the result of the baseline and $b$ is the result of the improved method.
\subsection{Results on Public and Industrial Datasets}
In the experiments, we set the learning rate to 0.001, embedding size to 32, batch size to 1024, and dropout rate to 0.5. The maximum length of the behavior sequence is set to 100 for all users. If the length of a user's behavior sequence is less than 100, we concatenate the behavior sequence with zero padding. In MAIL, dual autoencoders are used. The dimension of hidden layers in encoders is 1024-512-512, and the dimension of hidden layers in the decoders is 512-1024-$d_{t}$, where $d_{t}$ denotes the dimension of target vectors. The dimension of the hidden layers in the multilayer perceptron of the ranking tower is 512-256. The Adam optimizer \cite{31} is used with $\beta_{1} = 0.9$ and $\beta_{2} = 0.999$ in this paper. \par
The results are summarized in Table \uppercase\expandafter{\romannumeral2}. LLAE performs better than the basic EmbLR method. An AUC improvement of 0.6\% $\sim$ 0.9\% is obtained on new users, which is significant for large-scale real-world datasets. However, LLAE is limited to linear transformations and based on linear logistic regression. In comparison with the other five deep models, the weak representation ability of LLAE can be observed, which demonstrates the effectiveness of nonlinear transformation and a deep structure.\par
For MetaEmb and MAIL-Base, the BaseDNN model is used as a baseline and an AUC improvement of 0.8\% $\sim$ 1.3\% is obtained on new users. The proposed MAIL-Base generates behavior data based on user attributes to address CSR, while the MetaEmb gives a better initial embedding for the ranking model. In comparison, the generative strategy of MAIL is more effective than the initialization strategy of MetaEmb. Moreover, by comparing MAIL-Base and MAIL-DMR, it can be observed that MAIL can use better models in the ranking tower to achieve an incremental performance improvement. For the public dataset, MAIL improves the AUC on new users from 0.5934 to 0.5958 by replacing BaseDNN with DMR. For the industrial dataset, the AUC on new users is improved from 0.6849 to 0.6895. The model-agnostic property is very attractive for practical applications since there are various ranking models for different scenarios in the real world.\par

Furthermore, although cold-start methods, \ie, LLAE, MetaEmb, and MAIL, focus on the new users, we find that they show a slight AUC improvement on old users. For the public dataset, an AUC improvement of 0\% $\sim$ 0.06\% on old users can be observed. For the industrial dataset, an AUC improvement of 0\% $\sim$ 0.1\% on old users is obtained. This may be explained by noting that cold-start methods generate virtual behavior data or make a better initial embedding, which alleviates the sparse data problem for ranking models that enables performance improvement. The results of the GAUC metric are similar to those of AUC.
\begin{table}[thb]
\centering
    \textbf{\caption{Results of Ablation Study on Public and Industrial Datasets}}
    \vspace{-1em}
    \setlength{\tabcolsep}{3.5 mm}
    \centering
    \begin{tabular}{c|c|c|c|ccc}    
        \Xhline{1pt}
         \hline
         \multirow{2}*{\textbf{Metrics}}& \multicolumn{2}{c|}{New Users} & \multicolumn{2}{c}{Old Users} \\
         \cline{2-5}
         & AUC & RI  & AUC  & RI \\
         \hline
         \multicolumn{5}{c}{\textbf{Public Dataset}} \\
         \hline
         BaseDNN     & 0.5862 & -1.21\% & 0.6271 & 0.00\% \\
         MAIL-None   & 0.5834 & -1.68\% & 0.6263 & -0.12\% \\
         MAIL-Single & 0.5844 & -1.51\% & 0.6267 & -0.06 \\
         MAIL-Dual   & 0.5897 & -0.62\% & 0.6272 & 0.01\% \\
         MAIL-Base   & 0.5934 & 0.00\% & 0.6271 & 0.00\% \\
         \hline
         \multicolumn{5}{c}{\textbf{Industrial Dataset}} \\
        \hline
         BaseDNN     & 0.6771 & -1.13\% & 0.7788 & 0.00\% \\
         MAIL-None   & 0.6755 & -1.37\% & 0.7782 & -0.07\% \\
         MAIL-Single & 0.6764 & -1.24\% & 0.7784 & -0.05\% \\
         MAIL-Dual   & 0.6802 & -0.68\% & 0.7788 & 0.00\% \\
         MAIL-Base   & 0.6849 & 0.00\%  & 0.7788 & 0.00\% \\
        \Xhline{1pt}
        \end{tabular}
        \vspace{-1em}
\end{table}
\subsection{Ablation Study}
There are three items in the training loss $\mathcal{L}_{zst}$ of the zero-shot tower, including two cross-modal reconstruction items and an MMD item. Here, an ablation study is performed to explore the effectiveness of the three items. Five models are designed for comparison as follows:\par
1) \textbf{BaseDNN} BaseDNN is the ranking tower shown in Figure 2.\par
2) \textbf{MAIL-None} MAIL-None has no zero-shot training loss in the zero-shot tower, but only has a linear mapping from user attributes to user behaviors. The ranking tower of MAIL-None is BaseDNN. \par 
3) \textbf{MAIL-Single} The zero-shot tower of MAIL-Single is trained by Eq. (3). The ranking tower of MAIL-Single is BaseDNN.\par
4) \textbf{MAIL-Dual} The zero-shot tower of MAIL-Dual is trained by Eqs. (3) and (4). The ranking tower of MAIL-Dual is BaseDNN.\par
5) \textbf{MAIL-Base} The zero-shot tower of MAIL-Base is trained by Eq. (7). The ranking tower of MAIL-Dual is BaseDNN.\par
Table 2 gives the results of MAIL with different zero-shot towers on both public and industrial datasets. As the table indicates, MAIL-None and MAIL-Single show worse results than BaseDNN. For the public dataset, the AUC on new users of BaseDNN is 0.5862, while that of MAIL-None and MAIL-Single is 0.5834 and 0.5844, respectively. For the industrial dataset, the AUC on new users of BaseDNN is 0.6771, while that of MAIL-None and MAIL-Single is 0.6755 and 0.6764, respectively. The two models generate virtual data depending on user attributes and do not regularize the autoencoders by setting user behavior as input, which results in overfitting and performance degradation. In comparison, the results of MAIL-Dual and MAIL-Base are much better, the stronger regularization by dual cross-modal reconstruction and maximum mean discrepancy reduces the modal-shift problem between user behavior and user attributes to achieve satisfactory virtual data for the ranking tower.\par
Also, it is noticed that the unreliable data generated by MAIL-None and MAIL-Single slightly affect the model performance on old users. For the public dataset, an AUC decrease of 0.06\% $\sim$ 0.12\% is obtained. For the industrial dataset, an AUC decrease of 0.05\% $\sim$ 0.07\% is obtained. In comparison, MAIL-Dual and MAIL-Base are not observed to affect the ranking performance on old users.\par

\begin{figure}[!hbt]
\vspace{-0.3em}
\centering    
\subfigbottomskip = 1pt
\subfigcapskip = -5pt
\setlength{\abovecaptionskip}{0.1cm} 
\subfigure[The public dataset]{           
\includegraphics[width=0.42\textwidth]{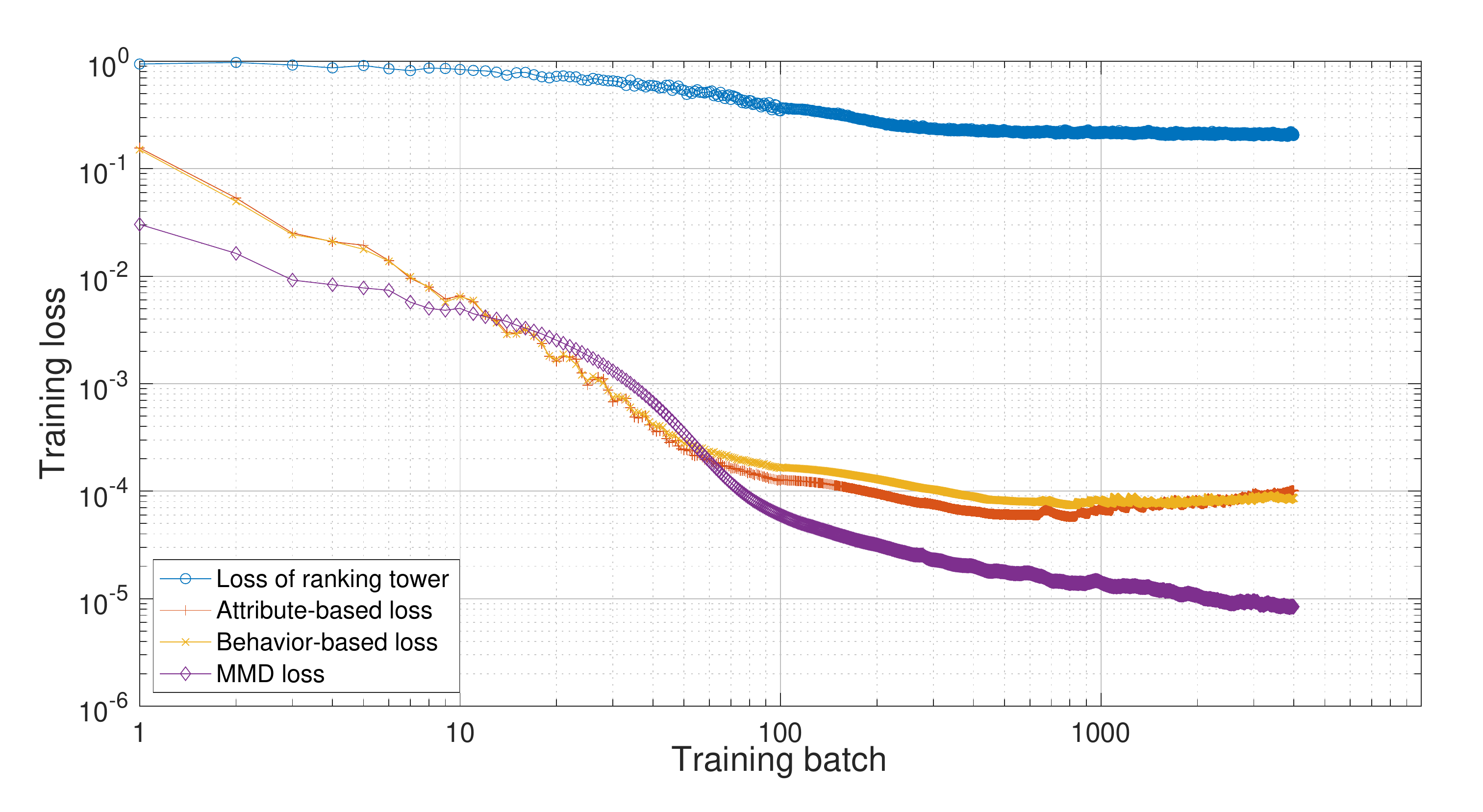}}
\subfigure[The industrial dataset]{
\includegraphics[width=0.42\textwidth]{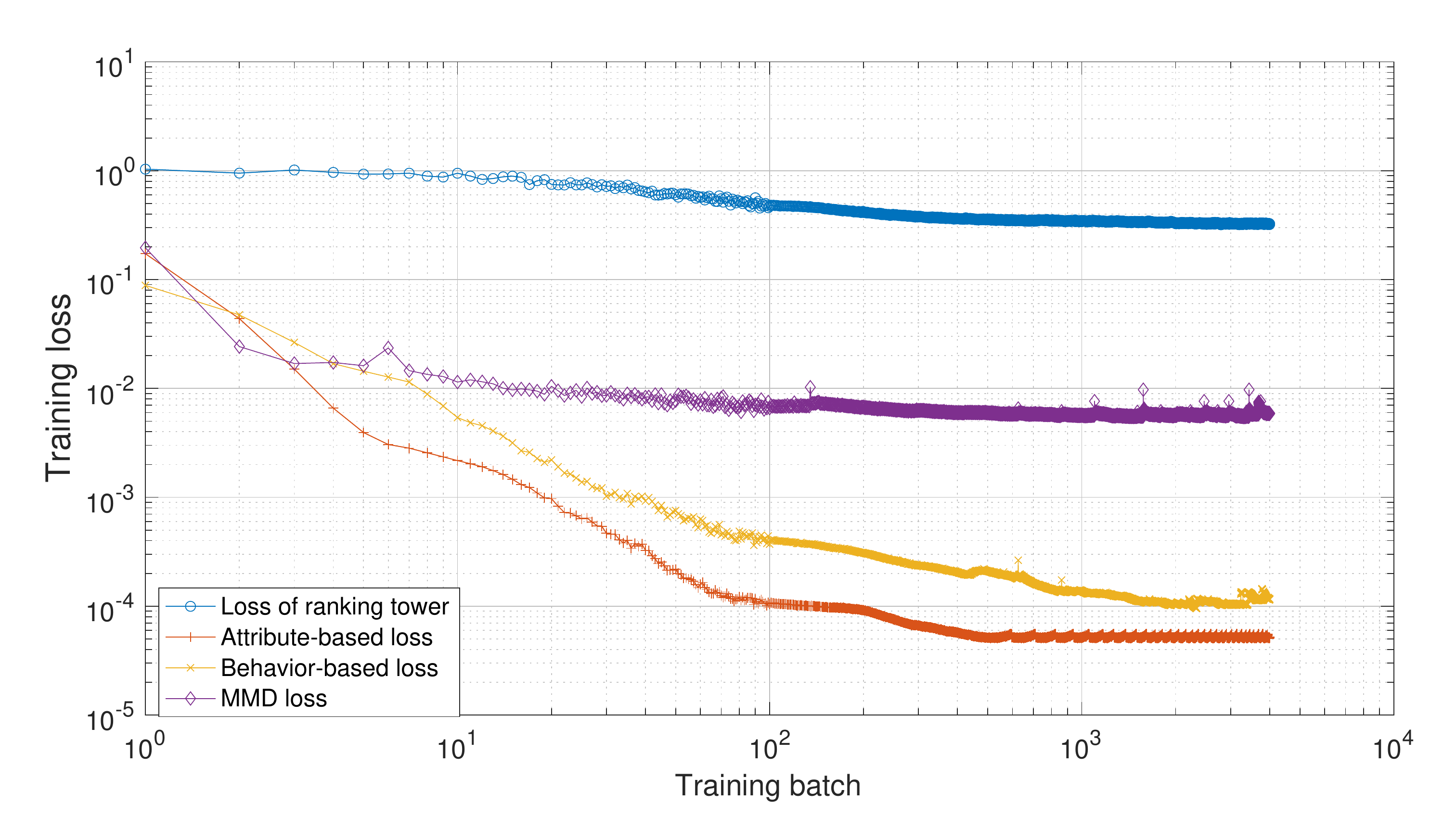}}
\caption{The training loss of MAIL on the public and industrial datasets.}
\end{figure}
In addition, we show the training loss of MAIL-Base in Figure 4, including the ranking tower loss $\mathcal{L}_{rt}$, the behavior-based cross-modal reconstruction loss $\mathcal{L}_{a}$, the attribute-based cross-modal reconstruction loss $\mathcal{L}_{v}$, and the MMD loss $\mathcal{L}_{d}$. Generally, the training loss decreases rapidly in one hundred batches for both of the public and industrial datasets, which verifies the learnability from user attributes to user behaviors. 
\begin{figure*}[!hbt]
\vspace{-0.3em}
\centering    
\subfigbottomskip = 1pt
\subfigcapskip = -5pt
\setlength{\abovecaptionskip}{0.1cm} 
\subfigure[Comparison between $\hat{\bm{v}}_{s}$ (yellow) and $\bm{v}_{s}$ (purple)]{           
\includegraphics[width=0.42\textwidth]{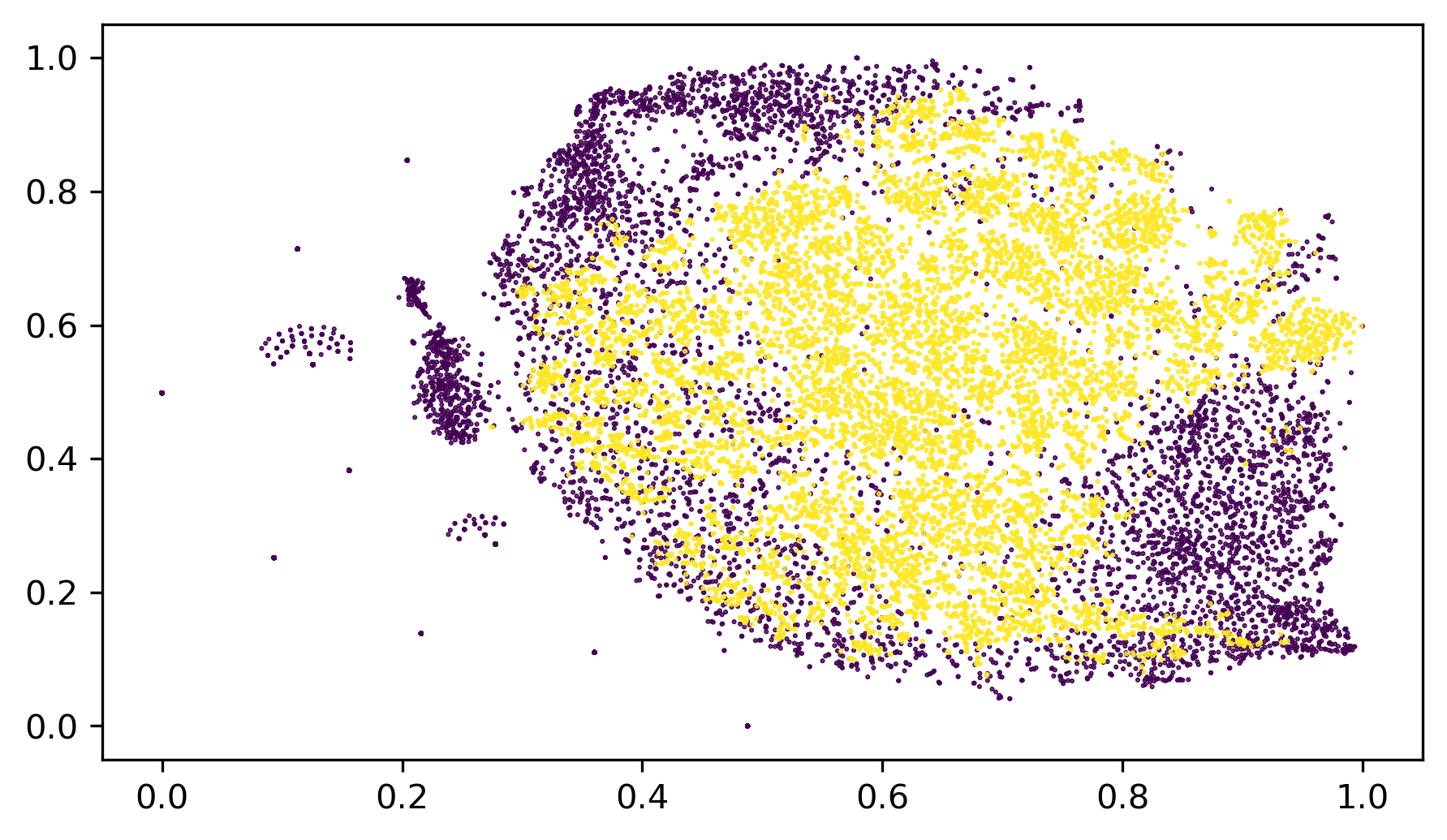}}
\subfigure[Comparison between $\hat{\bm{v}}_{o}$ (yellow) and $\bm{v}_{o}$ (purple)]{
\includegraphics[width=0.42\textwidth]{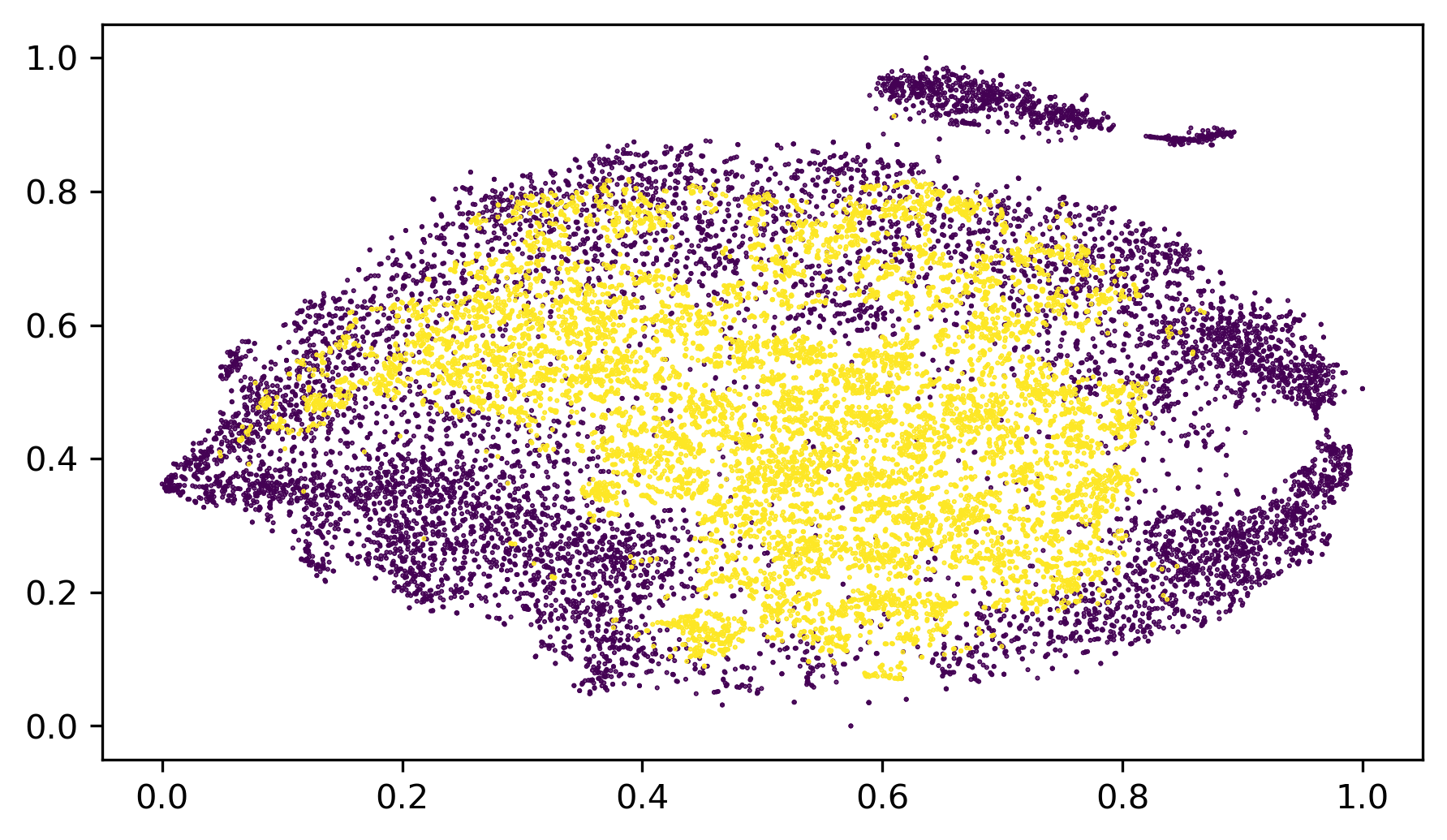}}
\caption{Visualization comparison on the public dataset.}
\vspace{-1em}
\end{figure*}

\begin{figure*}[t]
\vspace{-0.3em}
\centering    
\subfigbottomskip = 1pt
\subfigcapskip = -5pt
\setlength{\abovecaptionskip}{0.1cm} 
\subfigure[Comparison between $\bm{h}_{s}^{a}$ (yellow) and $\bm{h}_{o}^{v}$ (purple)]{           
\includegraphics[width=0.42\textwidth]{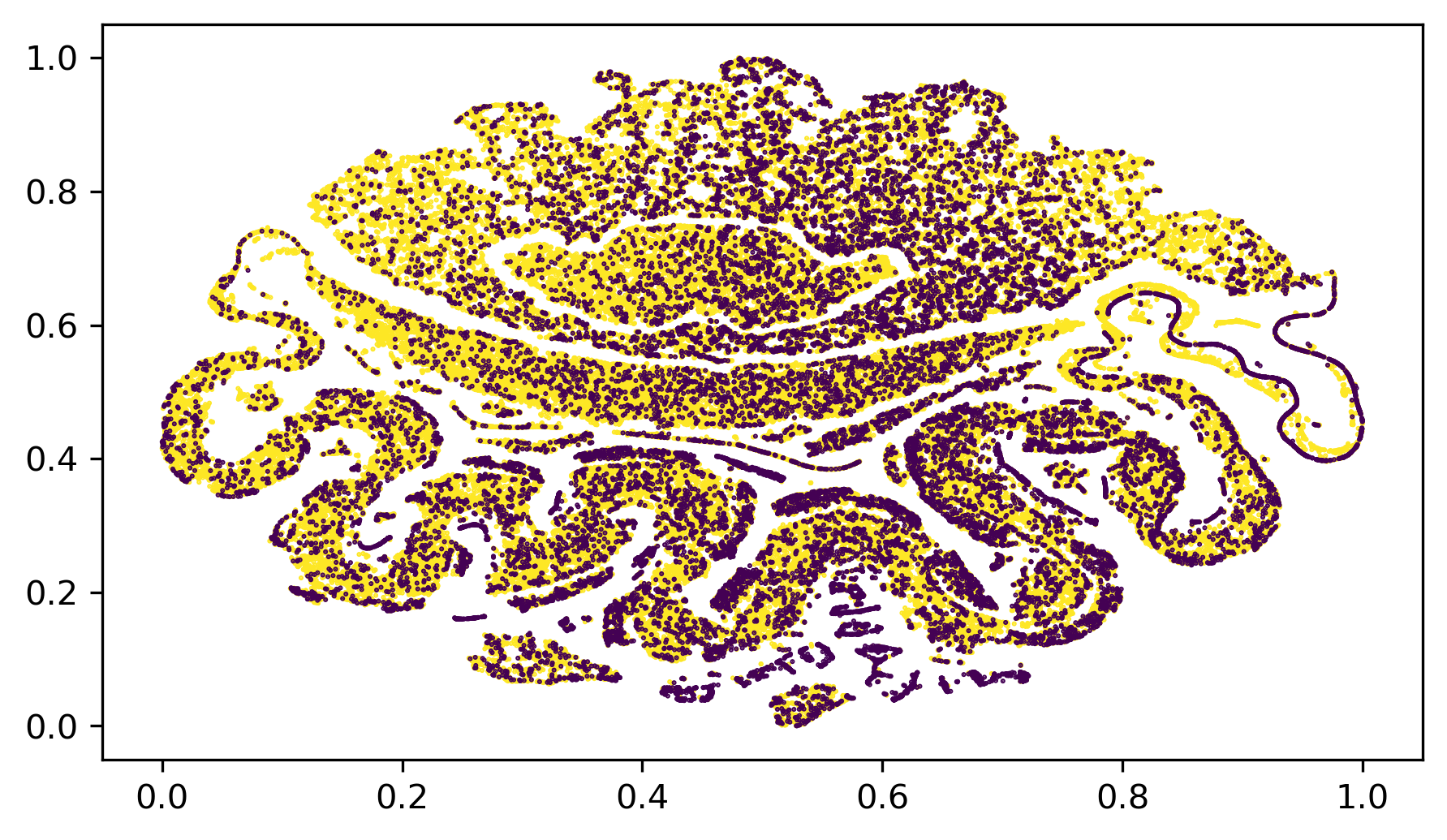}}
\subfigure[Comparison between $\hat{\bm{v}_{s}}$ (yellow) and $\hat{\bm{v}_{o}}$ (purple)]{
\includegraphics[width=0.42\textwidth]{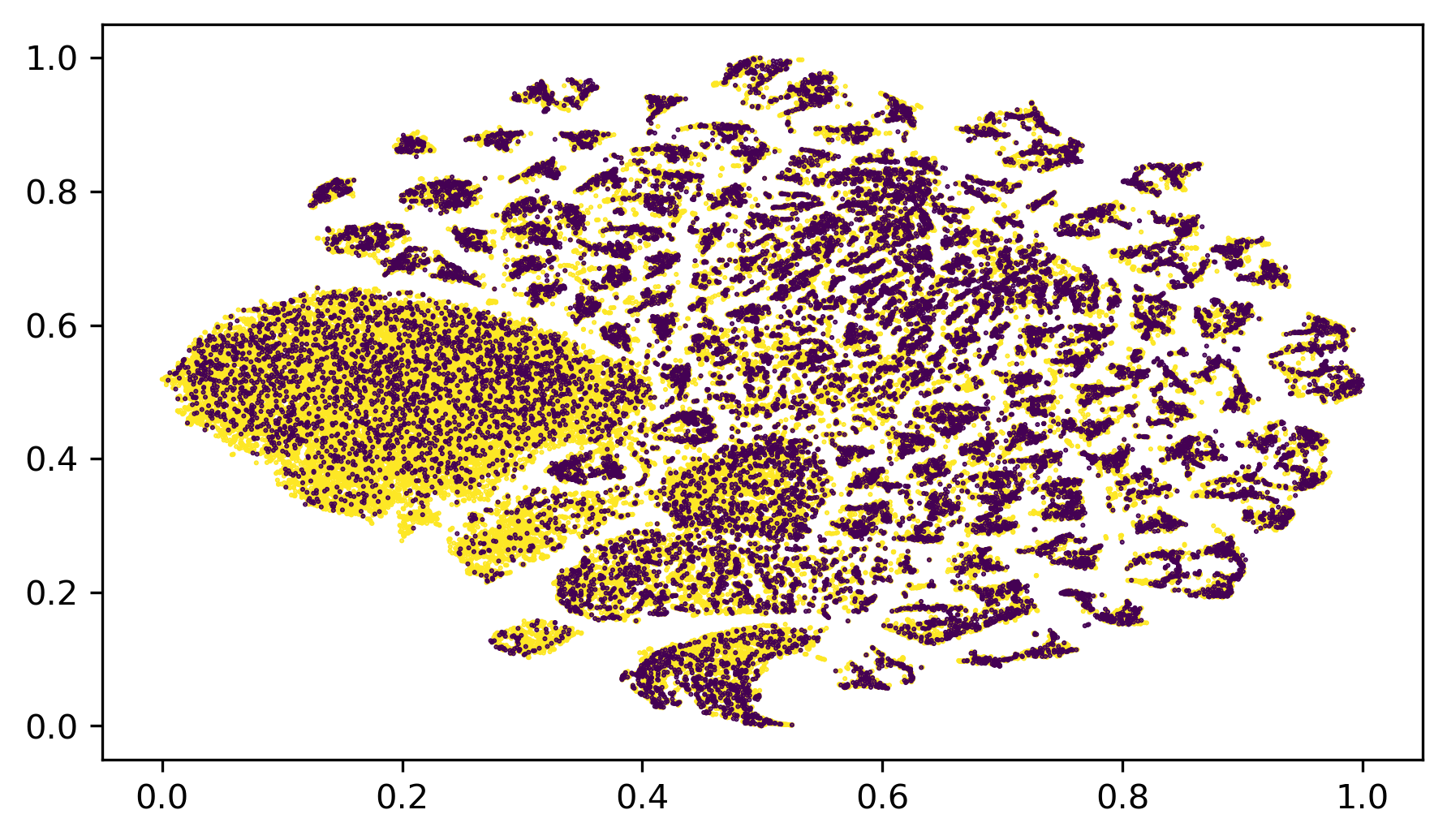}}
\caption{Visualization comparison on the industrial dataset.}
\end{figure*}
\subsection{Visualization Study}
In this subsection, we perform a visualization study for various hidden features in the zero-shot tower using the t-SNE \cite{36}.
\subsubsection{Visualization for the Public Dataset}
For the public dataset, we provide a visualization comparison between the generated features $\hat{\bm{v}}$ and the real features $\bm{v}$ to validate the effectiveness of the zero-shot tower. Note that the new users in the public dataset are made by removing their behavior data from the dataset; hence, we can perform the comparison on both the old users ($\hat{\bm{v}}_{s}$ and $\bm{v}_{s}$) and new users ($\hat{\bm{v}}_{o}$ and $\bm{v}_{o}$). The visualization is given in Figure 5. As shown, most of the yellow points are overlapped with purple points and the generated distribution is covered by the real distribution, which demonstrates the effectiveness of the zero-shot learning loss for the modal-shift problem.
\subsubsection{Visualization for the Industrial Dataset}
For the industrial dataset, we provide a visualization comparison between the new user's features and old user's features to validate the assumption that users enjoy a common attribute space to allow the behavior transfer from old users to new users. Specifically, $\bm{h}_{s}^{a}$ is compared with $\bm{h}_{o}^{v}$, and $\hat{\bm{v}_{s}}$ is compared with and $\hat{\bm{v}_{o}}$. The visualization is given in Figure 6. It is observed that the new users' attribute distribution $\bm{h}_{s}^{a}$ is almost the same as the old user's attribute distribution $\bm{h}_{o}^{v}$, which validates the assumption that all users enjoy a common attribute space. Moreover, based on the common feature space, the generated $\hat{\bm{v}_{s}}$ and $\hat{\bm{v}_{o}}$ are also similar to each other, which demonstrates the feasibility of attribute-based behavior transfer from old users to new users.
\subsection{Results from Online A/B Testing}
We conducted online A/B testing to validate the model performance for real applications in our recommender system. The MAIL-DMR and DMR were separately deployed for comparison, where three million users were assigned to each model. In the half-month of A/B testing, the proposed MAIL-DMR improves CTR by 13\% $\sim$ 15\% and CTCVR by 3\% $\sim$ 4\% relatively compared with DMR, which is the last version of the recommender model in our system. Note that users would perform some conversion actions, \eg, purchase goods and collect videos, only when they are genuinely interested in the recommended items. Hence, we highlight the contribution of MAIL-DMR to the improvement of CTCVR in real applications, which is more difficult than the improvement of CTR. 
\section{Conclusion}
In this paper, the MAIL model is proposed to address the cold-start problem for recommender systems in a two-tower framework. Inspired by the thought of zero-shot learning, a new zero-shot tower is designed and cotrained with a model-agnostic ranking tower in MAIL. By performing the cross-modal reconstruction and the maximum mean discrepancy-based modal alignment, the zero-shot tower generalizes behavior data from old users to new users based on their attributes. Using the virtual behavior data, the ranking tower can then capture new user's interests better for recommendation. An attractive property of the proposed method for practical application is that the ranking tower in MAIL can be implemented with any embedding-based ranking models, which allows MAIL to achieve an incremental performance improvement based on previous works. To verify the effectiveness of MAIL, ranking experiments, an ablation study, and a visualization study are performed on two large-scale real-world datasets. Furthermore, we have deployed the proposed MAIL on an online large-scale live recommendation system of a commercial application to show the significant improvements of CTR and CTCVR.


\bibliographystyle{ACM-Reference-Format}
\bibliography{sample-base}

\end{document}